\documentclass[twocolumn]{aastex631}

\newcommand{\spi}{{\it Spitzer}}
\newcommand{\her}{{\it Herschel}}
\newcommand{\galex}{{\it GALEX}}
\newcommand{\xmm}{{\it XMM-Newton}}

\newcommand{\kms}{km\,s$^{-1}$}

\newcommand\jwst{\emph{JWST}}

\newcommand\ngc{NGC 1808}

\shorttitle{Giant Biconical Dust Filaments in NGC~1808}
\shortauthors{Kane \& Veilleux}
\graphicspath{{./}{figures/}}

\begin{document}

\title{Giant Biconical Dust Filaments in the Starburst Galaxy NGC 1808}

\author[0009-0009-1074-3696]{Rohan Kane}
\affiliation{Department of Astronomy, University of Maryland, College Park, MD 20742, USA}

\author[0000-0002-3158-6820]{Sylvain Veilleux}
\affiliation{Department of Astronomy, University of Maryland, College Park, MD 20742, USA}
\affiliation{Joint Space-Science Institute, University of Maryland, College Park, MD 20742, USA}



\begin{abstract}
We present the results from an analysis of multi-wavelength archival data on 
the multi-phase outflow in the starburst galaxy \ngc. 
We report the detection at 70 and 100 \micron\ of dust filaments that extend up to $\sim$~13~kpc from the galactic mid-plane and trace an edge-brightened biconical structure along the minor axis of the galaxy. The inner filaments are roughly co-spatial with previously identified optical dust filaments, extraplanar polycyclic aromatic hydrocarbon emission, and neutral and ionized gaseous outflows.  The 70/160~\micron\ flux ratio, a proxy for dust temperature, is elevated along the edges of the cones, indicating that the dusty medium has been driven out of the central regions of these cones and possibly shock-heated by a large-scale outflow. We establish lower limits on the extraplanar dust mass and mean height above the stellar disk of $\log(M_d/{M_\odot})~=~6.48$ and $\vert\,z\,\vert~\sim~5$ kpc, respectively. The energy requirement of $(5.1-9.6)~\times~10^{56}$ ergs needed to lift the dusty material, assuming Milky-Way like dust-to-gas ratio, can be supplied by the current starburst, with measured star formation rate of $3.5-5.4~M_\odot$~yr$^{-1}$, over a timescale of $(4-26)$~$\xi^{-1}$~Myr, where $\xi$ is the efficiency of energy transfer. We conclude that a starburst-driven outflow is the most likely mechanism by which the dust features were formed. 
\end{abstract}


\section{Introduction} 
\label{sec:intro}

Dust is expected to move in and out of galaxies, tied to the inflow and outflow of gas. Gas enters galaxies through galaxy mergers and accretion flows, and flows out of galaxies driven by the energy from stellar processes and active galactic nuclei (AGN). Indeed, dust has been inferred to exist outside of galaxies in the gravitationally bound circumgalactic medium \cite[CGM, $\la$ 100-300 kpc;][]{Peek15}
and even the unbound intergalactic medium \citep[IGM; $\ga$ 100-300 kpc;][]{Mena12}. Studies of dust on such large scales rely on the statistical measurements of excess reddening of background sources. The importance of these large-scale flows of gas and dust on the evolution of galaxies is an area of active research \citep[][]{Tuml17,Veil20, Fauc23}. 

Direct detection of dust outside of galaxies remains sparse, especially beyond $\sim$ 10 kpc. Ultraviolet light emitted by young stars and reflected off extraplanar dust has been detected in several nearby edge-on star-forming and active disk galaxies \citep{Hoop05, Hodg14, Seon14, Hodg16, Shin18, Jo18}, but most of the dust observed in this fashion lies at vertical heights $\vert z \vert$ $\la$ 5 kpc from the galactic disk mid-plane. Brighter UV sources are needed to probe the dust beyond this distance using this method. UV-scattered ``cones" extending over 10-20 kpc along powerful gas outflows have been detected in a growing number of luminous quasars \citep{Zaka06, Obie16, Wyle16, Wyle22, Veil23}. 

A complementary method to study dust in and around galaxies is to search for the thermal radiation emitted by heated dust. Extraplanar mid-infrared (MIR) polycyclic aromatic hydrocarbon (PAH) emission out to $\vert z \vert$ $\approx$ 6 kpc has been reported in a few nearby edge-on starburst and active disk galaxies using the {\em Spitzer Space Telescope} \citep[\spi; e.g.,][] {Enge06, McCo13}. The {\em James Webb Space Telescope} \cite[\jwst;][]{Gard06, Rigb23} is rapidly building on the \spi\ legacy by providing exquisite MIR angular resolution and sensitivity to detect dust with $T \ga 100$ K. However, the dust tied to the more distant CGM, far from the heat sources in the host galaxy, is expected to be cooler and therefore more easily detectable in the far-infrared (FIR), beyond the reach of \jwst. The {\em Herschel Science Archive} ({\em HSA}\footnote{https://irsa.ipac.caltech.edu/applications/Herschel/}) of the {\em Herschel Space Observatory} \citep[\her;][]{Pilb10} has proven to be of great value to study the distribution and properties of the cool dust with temperatures $T \sim 10 - 100$ K in and around nearby galaxies. 
In particular, deep \her\ observations of edge-on disk galaxies have revealed extraplanar dust typically extending up to heights $\vert z \vert$ $\sim$ 3 kpc but sometimes up to 10-25 kpc \citep{Rous10, Mele15, McCo18, Yoon21, Veil21}.

The target of the present paper, \ngc, is a barred disk galaxy with inclination $i$ $\approx$ 57$^\circ$ \citep{Kori93} which has long been known to display conspicuous dust filaments that extend up to $\sim$ 3 kpc from the nucleus in optical broadband ($B - R$) color maps \citep{Phil93}. These filaments coincide with outflowing neutral-atomic gas filaments where the optical Na I D $\lambda\lambda$5890, 5896 doublet feature is seen blueshifted in absorption and redshifted in emission \citep{Phil93, Rupk21}. The inferred polar outflow is roughly along the minor axis of the stellar bar and large-scale disk (PA $\sim$ 45$^\circ$), where the northeast outflow lies in front of the disk and the southwest outflow is behind the disk \citep{Kori93, Shet10, Muno15}. The neutral-atomic gas has deprojected outflow velocities of up to $\sim$ 450 \kms\ \citep{Rupk21}. It is likely driven by the nuclear starburst in this system \citep{Forb92, Krab94, Junk95, Laur00, Heik07, Chen23}, although a low-luminosity, heavily obscured AGN, tentatively detected at high X-ray energies \citep{Jime05} and millimeter wavelengths \citep{Audi21}, may also contribute to the outflow energetics. This dusty neutral-atomic outflow roughly coincides spatially with the H~I 21-cm outflow reported by \citet{Kori93}, the dense cold molecular and neutral outflow studied by \citet{Sala16, Sala17, Sala18, Sala19} and \citet{Audi21}, the warm-ionized gas outflow observed in optical line emission \citep[e.g.,][]{Shar10, Rupk21}, the hot-ionized outflow detected in the soft X-rays with {\em ROSAT}, {\em Chandra}, and {\em XMM-Newton} \citep{Dahl94, Jime05}, and possibly also the base of a linearly polarized non-thermal radio continuum structure that extends up to 7 kpc from the nucleus \citep{Dahl90}. 

Here we report the results from a new analysis of the \her\ data on this object. These data are combined with the extensive set of ancillary data at other wavelengths 
to help us better understand our detection of cool-dust biconical filaments. Sec.\ \ref{sec:data} describes the multi-wavelength archival data used in our study. We present the results from our analysis of these data in Sec.\ \ref{sec:results} and discuss the implications of our results in Sec.\ \ref{sec:discussion}. The conclusions are summarized in Sec.\ \ref{sec:conclusions}. We adopt a distance of 12.3 Mpc for NGC~1808 \citep{Tull09, Tull13}, corresponding to a scale of 59 pc per arcsecond, for the remainder of this paper. 

\section{Archival Data} 
\label{sec:data}

A summary of the archival data used in the present study in given in Table \ref{tab:data}. 

\begin{table*}
\caption{Archival Data}
\label{tab:data}
\begin{center}
\begin{tabular}{cccccc}
\hline
Telescope & Instrument & PI & Dataset & Exposure (s) & Beam FWHM ($^{\prime\prime}$)\\
\hline
(1) & (2) & (3) & (4) & (5) & (6)\\
\hline\hline
\spi\ & IRAC 4.5 $\mu$m & D.\ Fisher & AOR 18284800 & 2072 & 1.78\\
\spi\ & IRAC 8 $\mu$m & D.\ Fisher & AOR 18284800 & 2072 & 3.16\\
\her\ & PACS 70 $\mu$m & E.\ Sturm & OBSID 1342204260 & 626 & 5.6\\
\her\ & PACS 100 $\mu$m & E.\ Sturm & OBSID 1342204262 & 626 & 6.8\\
\her\ & PACS 160 $\mu$m & E.\ Sturm & OBSID 1342204263 & 626 & 11.3\\
\her\ & SPIRE 250 $\mu$m & E.\ Sturm & OBSID 1342203633 & 1086 & 21.3\\
\her\ & SPIRE 350 $\mu$m & E.\ Sturm & OBSID 1342203633 & 1086 & 28.3\\
\her\ & SPIRE 500 $\mu$m & E.\ Sturm & OBSID 1342203633 & 1086 & 39.8\\
\galex\ & FUV Imager & N/A & NGS & 2769.2 & 4.0\\
\galex\ & NUV Imager & N/A & NGS & 7729.6 & 5.6\\
\xmm\ & Epic (pn, MOS1, MOS2) & F. Jansen & OBSID 0110980801 & 43606 & 6.0\\
\hline
\end{tabular}
\end{center}
\end{table*}

\subsection{\her}
\label{subsec:her}
We use archival observations of NGC 1808 taken with the \her\ Photodetector Array Camera and Spectrometer (PACS) at 70 \micron, 100 \micron, and 160 \micron\ under program ID KPGT\_esturm\_1. A pair of observations is taken initially at 70 \micron\ and 160 \micron\ followed by a pair of observations at 100 \micron\ and 160 {\micron\, so we co-add the 160 \micron\ maps. First, we background-subtract the images by averaging the medians of the flux measured within four elliptical areas far from the disk or extraplanar structures. 
We reproject all images to the pixel scale (1\farcs6) and units (Jy pixel$^{-1}$) of the 70 \micron\ image in a process that conserves the flux. The 100 \micron\ image does not need to be reprojected as its pixels already are 1\farcs6 and have units of Jy pixel$^{-1}$, but the 160 \micron\ image has 3\farcs2 pixels in the same units. The products are processed to Level 2.5 using the BULK\_REPROCESSING SPG v14.2.0 pipeline processing mode and mapped with {\tt JScanamorphos}. The 100 \micron\ image displays a slight gradient in which the southeastern background is brighter than the northwestern background.

A series of observations was also taken with the \her\ Spectral and Photometric Imaging Receiver (SPIRE) at 250 \micron, 350 \micron, and 500 \micron\ under KPGT\_esturm\_1. The respective pixel sizes of these images are 6\arcsec, 10\arcsec, and 14\arcsec. We reproject the background-subtracted SPIRE images to display with 1.6\arcsec\ pixels, while also converting from the native flux density units of Jy beam$^{-1}$ to Jy pixel$^{-1}$. The SPIRE products are processed to Level 2 with SPG v14.1.0 and mapped with {\tt JScanamorphos}.

\subsection{\spi}
\label{subsec:spi}
We use mid-infrared observations from the {\spi} archives taken with the InfraRed Array Camera (IRAC) at 4.5 \micron\ and 8 \micron\ under program ID 30496. Both images have a pixel size of 1\farcs22, and use native flux density units of MJy sr$^{-1}$. We clean horizontal streak artifacts in the 8 \micron\ image. 
We begin this process by selecting a region away from the disk that contains an artifact but extends slightly beyond the edges of the artifact into uncontaminated space.  We calculate the difference between the median of the entire region and the median of the sub-region affected by the artifact, and subtract that value from the artifact. 
Because of a large area of pixels that sharply drop off to negative values west of the galaxy at 8 {\micron}, we ignore the region 8 kpc west of the disk.}

\subsection{Ancillary Data}
We use ultraviolet observations taken with the far-ultraviolet (FUV) and near-ultraviolet (NUV) imagers on board the {\em Galaxy Evolution Explorer} ({\em GALEX}) as part of the Near Galaxy Survey (NGS) mission, with wavelength ranges 1350-1750 \AA\ and 1750-2800 \AA, respectively. The pixels in both bands are 1\farcs5. We background-subtract and reproject the \textit{GALEX} images to the 1\farcs6 pixel size of the 70 \micron\ image. Additionally, we use a series of X-ray images taken with all three \textit{XMM-Newton} European Photon Imaging Camera (EPIC) CCDs (MOS1, MOS2 \& pn)
. The \textit{XMM-Newton} images span five energy bandpasses ($0.2-2$, $2-4.5$, $4.5-7.5$, $7.5-10$, and $10-12$ keV), and have 4\arcsec\ pixels. 
The X-ray images are not background subtracted before reprojecting to the pixel size of the 70 \micron\ image. 

\section{Results}
\label{sec:results}

\subsection{Flux Maps}
\label{sec:fluxmaps}
\begin{figure*}
    \centering
    \includegraphics[width=\linewidth]{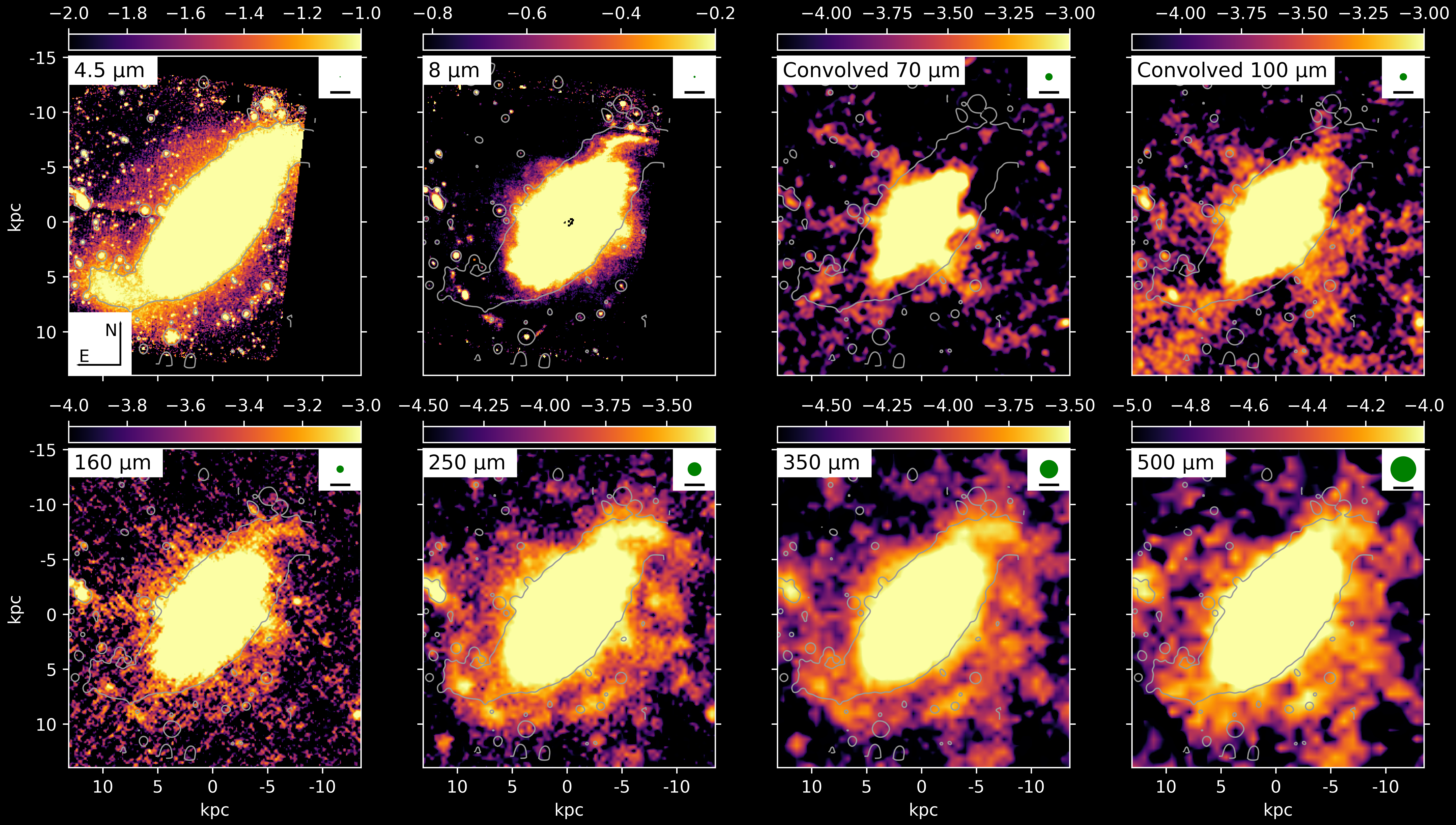}
    \caption{{\spi} and {\her} flux maps at their original resolution, with the exception of 70 and 100 \micron\ maps which have been convolved to the resolution of the 160 \micron\ map. Maps are in log scale. Each \spi\ pixel has units of MJy sr$^{-1}$. Each \her\ pixel has units of Jy pixel$^{-1}$. The color scale ranges from black (faintest flux) to yellow (brightest flux). The gray contour shown in each panel illustrates the extent of the stellar disk as defined at 4.5 \micron\ (0.075 MJy sr$^{-1}$). Beam sizes are represented by the green circle in the top right of each panel. The black line underneath is 2 kpc.  North is at the top and east to the left.}
    \label{fig:FluxMaps}
\end{figure*}

Both {\spi} images in Fig.\ \ref{fig:FluxMaps} show the characteristic barred spiral structure of {\ngc}. At 4.5 \micron, the emission largely traces the stellar continuum. We use the method outlined in \citet{McCo13} to derive a characteristic scale height, assuming the flux intensity decreases exponentially from the mid-line of the stellar bar (disk). For each pixel along the mid-line, we measure the distance from the mid-line until the relationship $I(z_0) = I_0e^{-1}$ is satisfied, where $z_0$ is the distance from the mid-line, I($z_0$) is the intensity at $z_0$, and I$_0$ is the intensity of the pixel in the mid-line. The scale height at that position along the mid-line is taken as the average of $z_0$ values measured on either side of the mid-line. We derive a scale height for the entire bar (disk) by averaging the scale height at each pixel along the mid-line, yielding a value of $z_0 = 1.3$ kpc, after correction for the inclination of the disk ($i = 57^{\circ}$). 

In Fig.\ \ref{fig:FluxMaps}, we also show one of the isophotal contours of the bar where the 4.5 \micron\ flux, after reprojection to the pixel scale of the 70 \micron\ map and convolution to the resolution of the 160 \micron\ map, is equal to 0.075 MJy sr$^{-1}$, a value 
we chose for its ability to indicate where the spiral arms  emerge from the bar without 
including fainter extraplanar emission.

\begin{figure*}
    \centering
    \includegraphics[width=\linewidth]{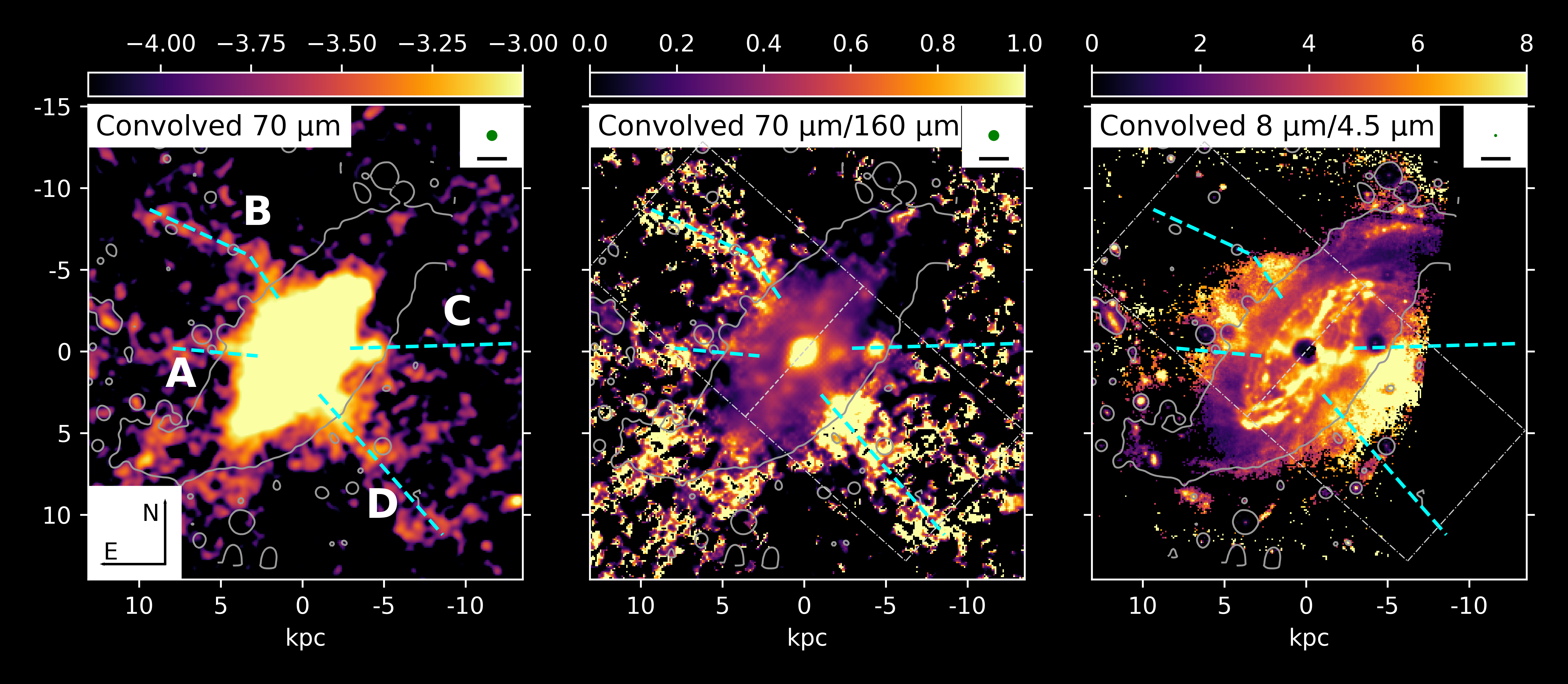}
    \caption{(From Left to Right) The 70 \micron\ flux map convolved to the resolution of the 160 \micron\ flux map, the 70-to-160 \micron\ flux ratio map, 
    and the 8-to-4.5 \micron\ flux ratio map, respectively. The 70 \micron\ flux map is in log scale, where each pixel has units of Jy pixel$^{-1}$. The cyan dashed lines trace the extended filamentary emission detected at 70 and 100 \micron. The dashed rectangular boxes in the middle and right panels indicate where the vertical profiles of the flux ratios presented in Fig.\ \ref{fig:ratioprofile} are measured. The color scale ranges from black (lowest values) to yellow (highest values). The gray contour shown in each panel illustrates the extent of the stellar disk as defined at 4.5 \micron\ (0.075 MJy sr$^{-1}$). Beam sizes are represented by the green circle in the top right of each panel. The black line underneath is 2 kpc. North is at the top and east to the left.}
    \label{fig:lines}
\end{figure*}

At 8 \micron, the emission from polycyclic aromatic hydrocarbons (PAHs) dominates and remains elevated at $\vert z \vert$ heights well beyond $z_0$, where the 4.5 \micron\ stellar component has mostly faded into the background (Fig.\ \ref{fig:FluxMaps}). We return to this result in Sec. \ref{sec:FluxRatioMaps}, where we present the 8-to-4.5 \micron\ flux ratio map to illustrate this point. 
The \her\ 70 \micron\ and 100 \micron\ maps in Fig.\ \ref{fig:FluxMaps} show emission both in the northeastern extraplanar region analyzed by \citet{Phil93} and southwest of the bar structure. The 70 \micron\ map shows filamentary structures that extend out of the bar region
(along lines A, B, C \& D in Fig. \ref{fig:lines}), tracing a pattern that suggests the edges of a bicone. Filament B on the northern side extends out to a height of up to $\sim$ 13 kpc from the disk mid-plane. Here and throughout the text, we do not correct any height measurements for projection effects as the locations of these filaments with respect to the plane of the sky are uncertain. The two filaments in the southern extraplanar region are well-defined to a height of $\sim 1-2$ kpc above the disk, and then the emission becomes patchy out to $\sim$9 kpc from the bar, reaching the far southwestern edge of the image. The 100 \micron\ map shows similar features.  
At 100 \micron\ there is an apparent excess of flux in the southeastern corner of the image which likely is due to the slight positive gradient from NW to SE of the background intensity (discussed in Sec. \ref{subsec:her}). Part of this emission may also be due to cirrus-like emission in the disk of the host galaxy itself.
The southernmost filament (line D) is not visible at 160 \micron, and the overall flux distribution is less filamentary and more patchy. 
At longer wavelengths, in the SPIRE maps (Fig.\ \ref{fig:FluxMaps}), the filaments disappear and the structure of the spiral arms starts to emerge.

\subsection{Flux Ratio Maps}
\label{sec:FluxRatioMaps}

We use the \her\ PACS images at 70 \micron\ and 160 \micron\ to map the structure of the heated extraplanar dust. As FIR emission in galaxies often peaks between these wavelengths \citep{Mele15}, the 70-to-160 \micron\ flux ratio, $S_{70}/S_{160}$, is a good indicator of the dust temperature, assuming the other properties of the dust (e.g., composition, grain size distribution) remain constant. For instance, this ratio is equal to 0.2 
for a modified blackbody at $T \sim\ 20$ K, 0.5 at $T \sim\ 24$ K, and 1 at $T \sim\ 28$ K. With this in mind, we convolve the 70 \micron\ map to the resolution of the 160 \micron\ data (11\farcs3) and take the ratio of the convolved 70 \micron\ map to the 160 \micron\ map (Fig.\ \ref{fig:lines}). 
Some of the highest flux ratios are found along the northernmost and southernmost filaments (lines B \& D).
In the southern cone, there is a distinct knot of high 70-to-160 \micron\ flux ratios near $\sim$ 6 kpc from the galactic center and a second knot further south at a distance $\sim$ 8 kpc. While not as high as along the filaments, the values of the flux ratios in the region between the filaments are also  higher than the values in the disk. Some of the patchy emission seen in the southwestern corner of the 70 \micron\ map does not appear in the ratio map because the 160 \micron\ emission was often undetected in these regions.

Next, we use the ratio of the {\spi} IRAC images at 8 \micron\ and 4.5 \micron\ to map the PAH emission relative to that of the stellar continuum (Fig.\ \ref{fig:lines}). The $S_8/S_{4.5}$ ratio map shows a distinct $\Theta$-like structure where the ratios are generally higher in the central region and within the inner northeast and southwest conical regions traced by the filaments at 70 and 100 \micron. The dip in the center of this map is an artifact caused by saturation in the center of the 8 \micron\ map.

\begin{figure}
    \centering
    \includegraphics[width=\columnwidth]{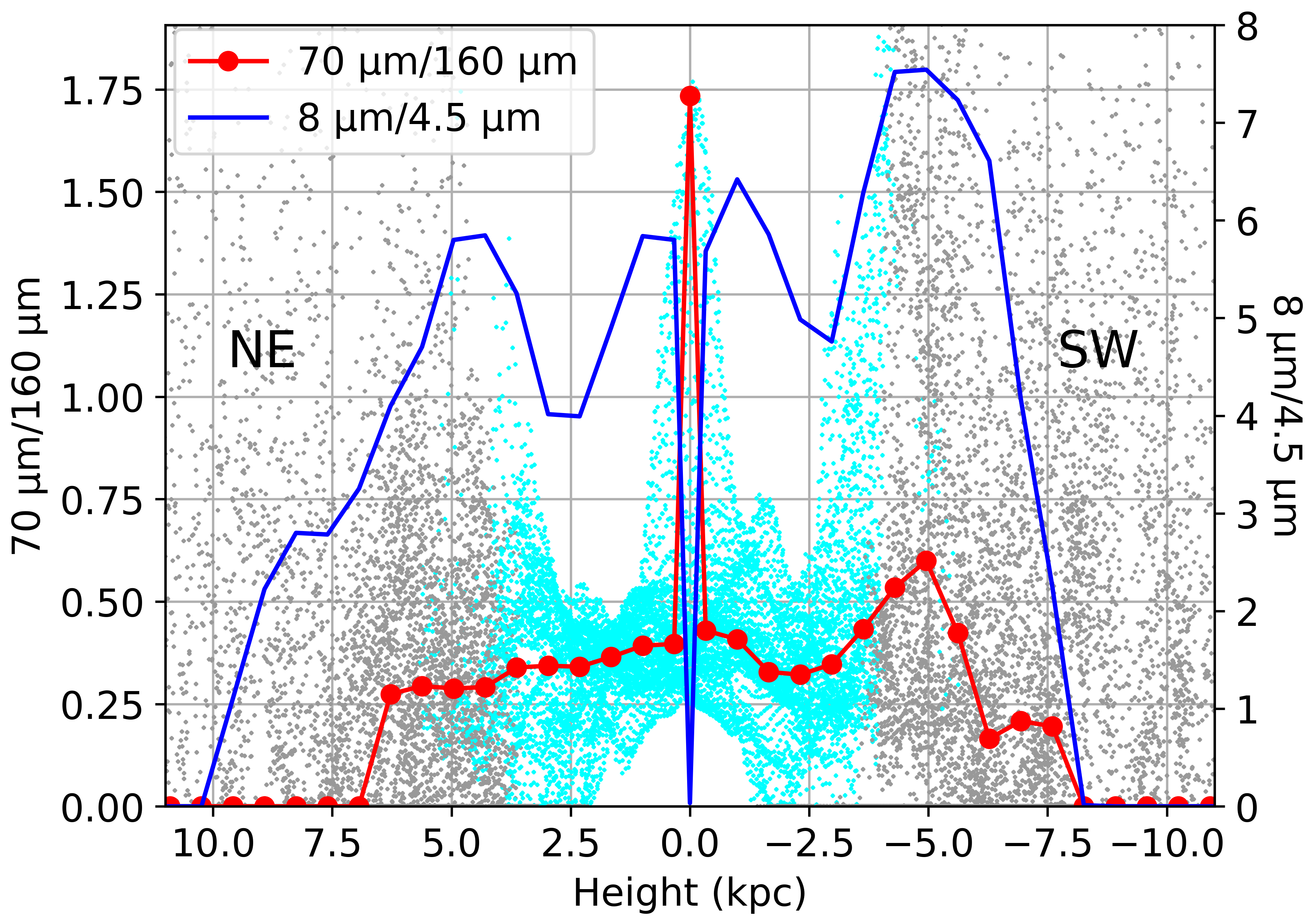}
    \caption{Vertical profiles of the $S_{70}/S_{160}$ (left axis) and $S_{8}/S_{4.5}$ (right axis) flux ratios along the minor axis of the disk (within the rectangular boxes shown in Fig.\ \ref{fig:lines}). The individual values of $S_{70}/S_{160}$ within these rectangular boxes are shown in gray, the medians are connected by a solid line (red = $S_{70}/S_{160}$, blue = $S_{8}/S_{4.5}$). Cyan points are within the stellar bar (4.5 \micron\ flux $\ge$ 0.075 MJy sr$^{-1}$). The central dip in the $S_{8}/S_{4.5}$ profile is an artifact due to saturation of the 8 \micron\ map in the nucleus.} 
    \label{fig:ratioprofile}
\end{figure}

Fig. \ref{fig:ratioprofile} shows both the $S_{70}/S_{160}$ and $S_{8}/S_{4.5}$ ratios as a function of height from the galactic mid-plane. The FIR ratios in the southern dust feature peak just outside the edge of the stellar bar, before reaching a local minimum between $6-7$ kpc. Although the $S_{70}/S_{160}$ ratios begin to rise again beyond 7 kpc, this next local peak in the ratios occurs in conjunction with the spiral arms making it difficult to distinguish between extraplanar emission and the underlying disk emission, so we adopt a conservative maximum emission height of 7 kpc for the southwestern dust feature. Similarly, due to the drop-off in both $S_{70}/S_{160}$ and $S_{8}/S_{4.5}$ ratios beyond 7 kpc in the northeast region, we adopt 7 kpc as a conservative maximum height for the northeastern dust feature.

\subsection{Comparison to UV and X-Ray Images}

\begin{figure*}
    \centering
    \includegraphics[width=\linewidth]{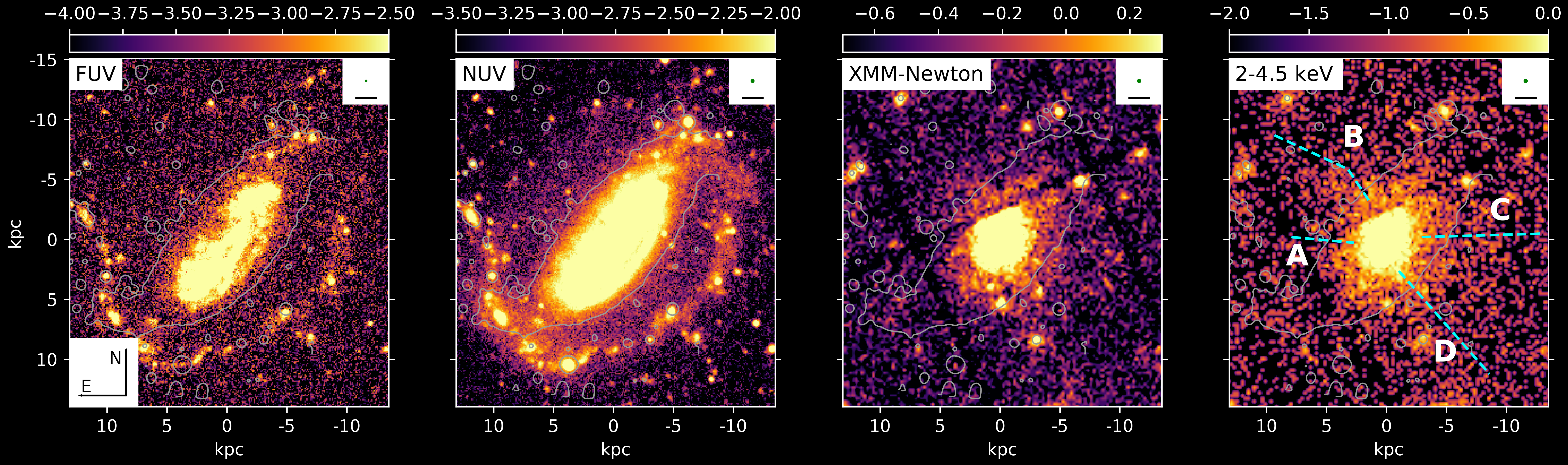}
    \caption{: (From left to right) {\it GALEX} FUV and NUV flux maps, Composite ($0.2-12$ keV) {\it XMM-Newton} flux map, and $2-4.5$ keV filter flux map. Maps are in log scale. UV maps have units of counts sec$^{-1}$, X-Ray maps have units of counts. The cyan dashed lines trace the extended filamentary emission detected at 70 and 100 \micron. The color scale ranges from black (faintest) to yellow (brightest). The gray contour shown in each panel illustrates the extent of the stellar disk as defined at 4.5 \micron\ (0.075 MJy sr$^{-1}$). Beam sizes are represented by the green circle in the top right of each panel. The black line underneath is 2 kpc. North is at the top and east to the left.}
    \label{fig:UVXRay}
\end{figure*}

While the spiral structure outside the central bar of \ngc\ is prominent in both {\em GALEX} FUV and NUV images (Fig.\ \ref{fig:UVXRay}), the UV emission is notably fainter in the northeast quadrant, possibly due to obscuration by foreground dust associated with the FIR filaments. Part of the spiral structure in the southwest quadrant is co-spatial with the local maximum in the $S_{70}/S_{160}$ ratio at $\sim$ 8 kpc southwest from the nucleus noted in the previous section.  This possible contamination of the extraplanar FIR emission by the spiral arms will be taken into account when deriving the properties of the dusty extraplanar filaments (Sec.\ \ref{sec:masses_temperatures}). 

The extraplanar X-ray emission in the \textit{XMM-Newton} data, first pointed out in \citet{Jime05}, is most visible at $2-4.5$ keV (Fig.\ \ref{fig:UVXRay}), where a streak of emission is aligned with the northernmost feature seen at 70 \micron\ (line B). However, across the full energy band of \textit{XMM-Newton} (third panel of Fig.\ \ref{fig:UVXRay}), the X-ray emission is largely associated with the disk, including some point sources within the spiral arms. 

\subsection{Dust Masses and Temperatures}
\label{sec:masses_temperatures}
\begin{figure}
    \centering
    \includegraphics[width=0.9\columnwidth]{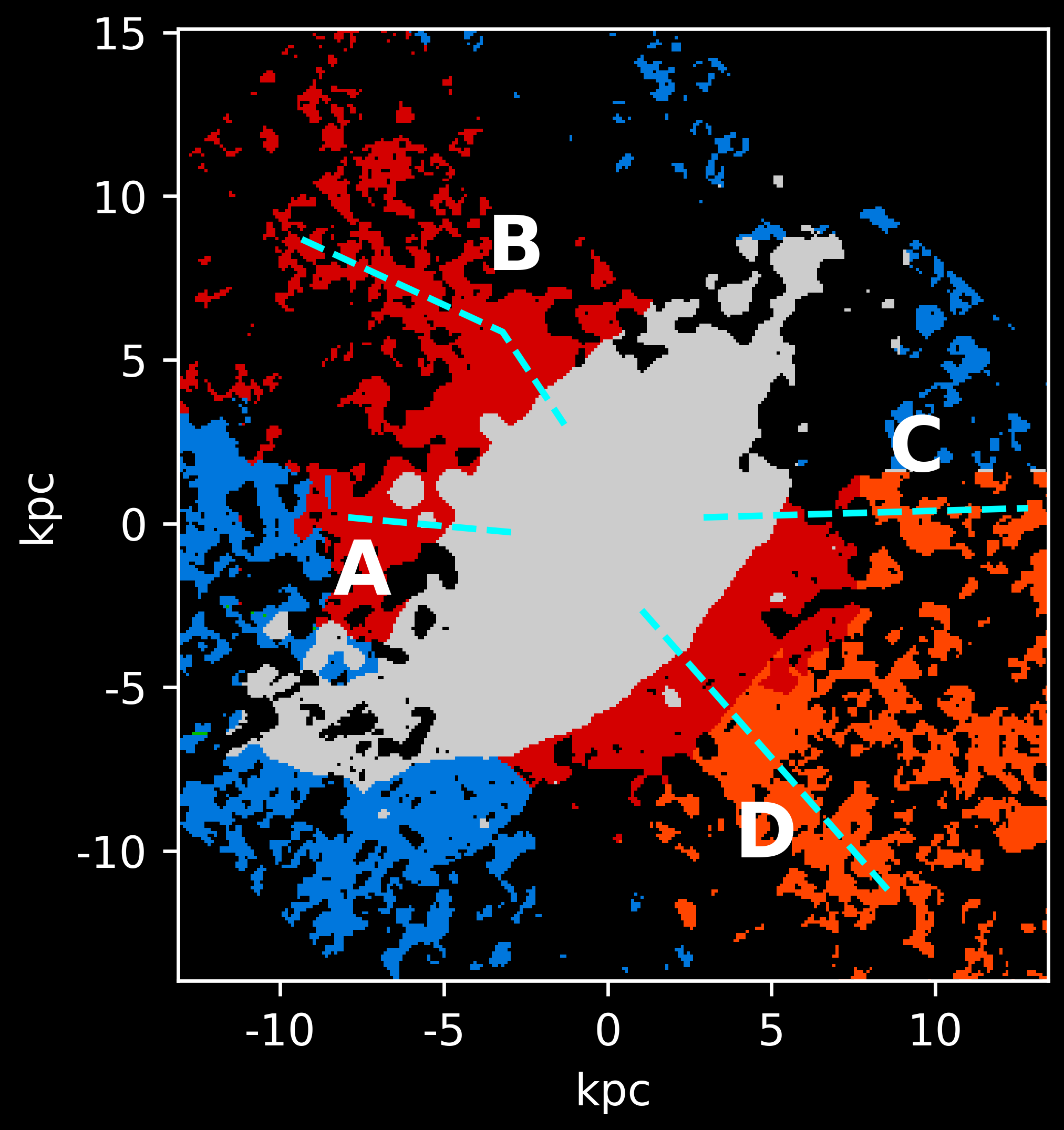}
    \caption{Regions used to estimate the dust masses and temperatures in \ngc. The stellar bar region is shown in gray, the spiral arms (excluded from the analysis) are shown in blue, and the northern and southern cones are shown in red. The southern cone has been displayed in two shades of red to delineate the conservative region (shown in dark red) uncontaminated with underlying emission from the spiral arms (shown in light red). 
    The cyan dashed lines mark the ridges of the 70 \micron\ filaments from Fig.\ \ref{fig:lines}. North is up and east is left.}
    \label{fig:subregions}
\end{figure}

In order to estimate dust masses and temperatures, we divide the galaxy into a series of regions (Fig.\ \ref{fig:subregions}), including the stellar bar region, the northern extraplanar region, and the southern extraplanar region. The stellar bar is shown in gray, extraplanar regions are shown in shades of red, and unanalyzed regions of the spiral arms are shown in blue. Since the emission in the southern extraplanar region is contaminated by the prominent spiral structure in the foreground disk, we split the southern extraplanar region into two, one that captures the space between the 4.5 \micron\ stellar bar and UV spiral arms (shown in darker red), and another one that includes the potentially contaminated emission (shown in a lighter red). The cone edges labeled C and D extend from the uncontaminated (dark red) region into potentially contaminated space (light red), while the cone edges labeled A and B are not affected by contamination. We again define the stellar bar region by selecting pixels above 0.075 MJy sr$^{-1}$ in the 4.5 \micron\ map (after reprojection to the 70 \micron\ map scale and convolution to the 160 \micron\ map beam size), a value we found to both trace the emergence of spiral arms from the stellar bar and avoid including
the extraplanar regions. The fluxes extracted from these regions are listed in Table \ref{tab:fluxes}.

We derive the dust masses and temperatures in these regions by fitting a single-temperature modified blackbody distribution to the spectral energy distribution (SED) measured from the \her\ PACS and SPIRE images according to 
\begin{equation}
    F_\nu = \frac{M_d\kappa_{\nu}B_\nu(T_d)} {D^2}
\end{equation}
where $F_\nu$ is the flux at the associated wavelength, $M_d$ is the fitted dust mass, $\kappa_\nu$ is the dust opacity at the associated wavelength, and $B_{\nu}(T_{d})$ is the Planck function of the fitted temperature. The dust opacity, $\kappa_\nu = \kappa_{0}(\nu/\nu_{0})^{\beta}$, is calculated based on a value of $\kappa_0 = 0.192$ m$^{2}$ kg$^{-1}$ at 350 \micron, and holding dust emissivity as a fixed parameter at $\beta = 2.0$ \citep{Mele15}. The IR spectrum above 60 \micron\ is dominated by steady thermal emission from large dust grains as opposed to the time-dependence seen in smaller grains \citep{Drai03}, which dominates the emission below this wavelength. So, to 
reduce the uncertainty stemming from small-grain fluctuations, we follow \citet{Mele15} and fit the FIR fluxes holding the 70 \micron\ flux as an upper limit. To assign uncertainties to the integrated fluxes in all other bands, we add in quadrature the error on mean brightness in each pixel, the standard deviation of pixels used to take the background, and a calibration uncertainty. This calibration uncertainty is 20\% in the bar region, increased to 40\% in the extraplanar regions, and 50\% for the 100 \micron\ measurements in the northern region to account for the gradient in the background. The uncertainties on the mass and temperature from the fits are calculated by taking the 1-$\sigma$ spread of 100 SED fits calculated from random photometric data points within the observed error bars. 

\begin{figure}
    \centering
    \includegraphics[width=\columnwidth]{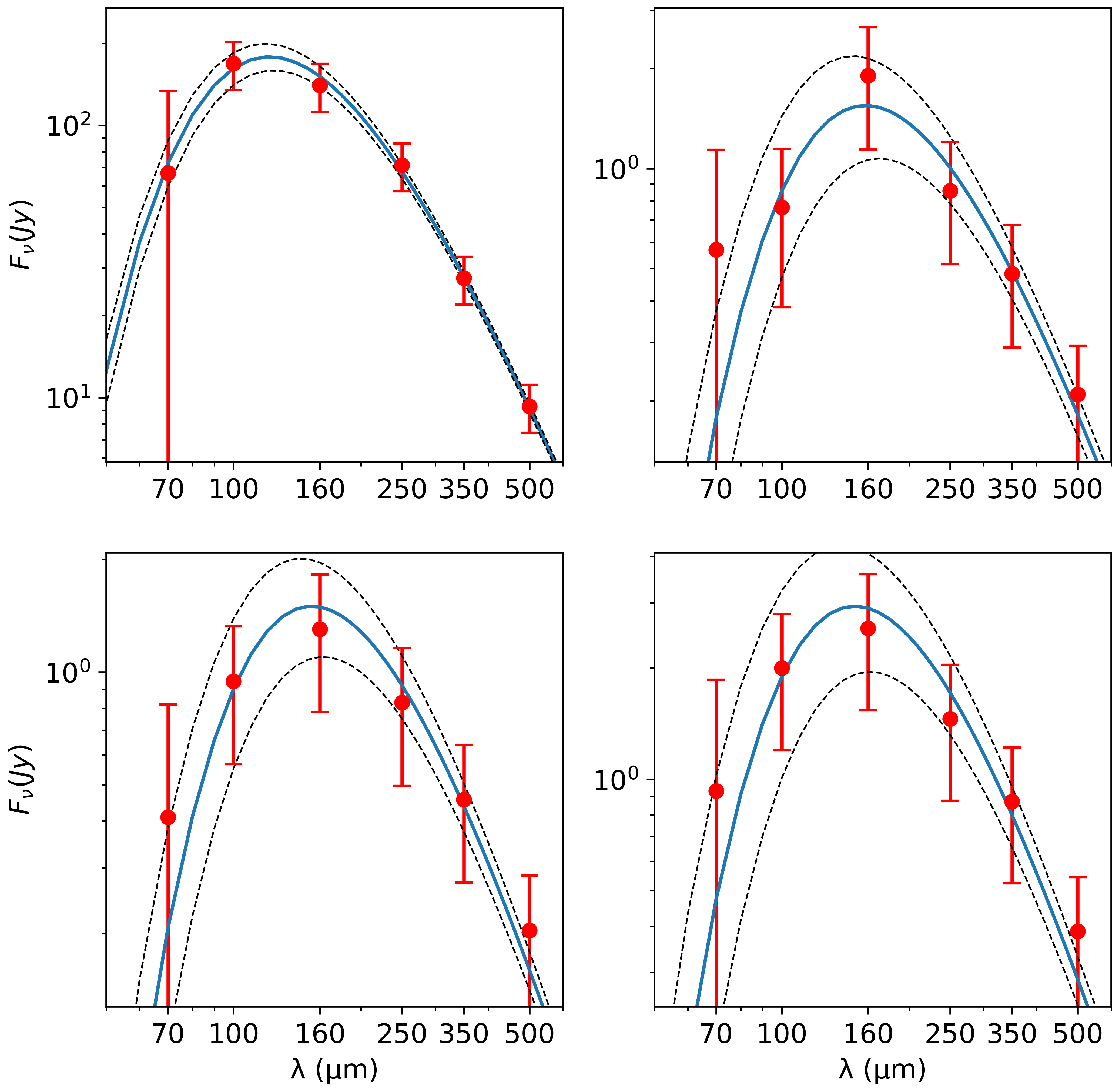}
    \caption{SED fitting. \textit{Top Left}: Stellar bar (gray region in Fig.\ \ref{fig:subregions}). \textit{Top Right}: Northern outflow (top left red region in Fig. \ref{fig:subregions}. \textit{Bottom Left}: Southern outflow between stellar bar and spiral arms (bottom right dark-red region in Fig.\ \ref{fig:subregions}). \textit{Bottom Right}: Southern outflow, including region with the spiral arms (bottom right light-red region in Fig.\ \ref{fig:subregions}). Masses and temperatures derived from these curves are listed in Table \ref{tab:dust}.}
    \label{fig:SEDs}
\end{figure}

The results from the SED fitting are shown in Fig.\ \ref{fig:SEDs} and summarized in Table \ref{tab:dust}. We find that the dust in the cones is overall slightly cooler than the dust in the stellar bar, despite the 70-to-160 \micron\ ratio map indicating the filaments along the edges are warmer than the disk. Extraplanar dust constitutes 5\% of the total dust mass in the galaxy. This value increases to 7\% if we include the dust in the southern cone which may be contaminated by spiral arm emission. 

\begin{table*}
\begin{caption}{Integrated Far-Infrared Fluxes (in Jy)}
\label{tab:fluxes}
\end{caption}
\begin{center}
\begin{tabular}{c c c c c}
\hline
Band $\lambda$ & Stellar Bar & N Cone & S Cone (Min.)$^{(a)}$ & S Cone (Max.)$^{(a)}$ \\
\hline
(1) & (2) & (3) & (4) & (5) \\
\hline\hline
PACS 70 \micron\ & 67.0 $\pm$ 67.0 & 0.6 $\pm$ 0.6 & 0.4 $\pm$ 0.4 & 0.9 $\pm$ 0.9 \\
PACS 100 \micron\ & 168.9 $\pm$ 33.8 & 0.8 $\pm$ 0.4 & 1.0 $\pm$ 0.4 & 2.0 $\pm$ 0.8 \\
PACS 160 \micron\ & 140.3 $\pm$ 28.1 & 1.9 $\pm$ 0.8 & 1.3 $\pm$ 0.5 & 2.6 $\pm$ 1.0 \\
SPIRE 250 \micron\ & 71.6 $\pm$ 14.3 & 0.9 $\pm$ 0.3 & 0.8 $\pm$ 0.3 & 1.5 $\pm$ 0.6 \\
SPIRE 350 \micron\ & 27.5 $\pm$ 5.5 & 0.5 $\pm$ 0.2 & 0.5 $\pm$ 0.2 & 0.9 $\pm$ 0.4 \\
SPIRE 500 \micron\ & 9.3 $\pm$ 1.9 & 0.2 $\pm$ 0.1 & 0.2 $\pm$ 0.1 & 0.4 $\pm$ 0.2 \\
\hline
\end{tabular}
\end{center}
$^{(a)}$ The region labeled ``South Cone (Min.)'' is the sub-region indicated by the darker shade of red in Fig.\ \ref{fig:subregions}. The region labeled ``South Cone (Max.)'' is the entire southwestern region including the lighter shade of red in this same figure.
\end{table*}

\begin{table}
\begin{caption}{Dust Masses and Temperatures}
\label{tab:dust}
\end{caption}
\begin{center}
\begin{tabular}{c c c}
\hline\hline
Region & Mass & Temp. \\
& $\log(M_d/{M_\odot})$ & ($K$) \\
\hline
(1) & (2) & (3) \\
\hline
Stellar Bar & 7.72$^{+0.03}_{-0.03}$ & 23.9 $\pm$ 0.5 \\
North Cone & 6.22$^{+0.10}_{-0.13}$ & 18.4 $\pm$ 1.3 \\
South Cone (Min.)$^{(a)}$ & 6.14$^{+0.10}_{-0.13}$ & 19.0 $\pm$ 1.1 \\
South Cone (Max.)$^{(a)}$ & 6.37$^{+0.14}_{-0.21}$ & 19.5 $\pm$ 1.5 \\
\hline
\end{tabular}
\end{center}
$^{(a)}$ The region labeled ``South Cone (Min.)'' is the sub-region indicated by the darker shade of red in Fig.\ \ref{fig:subregions}. The region labeled ``South Cone (Max.)'' is the entire southwestern region including the lighter shade of red in this same figure. 
\end{table}

\subsection{Star Formation Rate in the Stellar Bar}
\label{sec:SFR}

First, we estimate the IR-based star formation rate (SFR) in the stellar bar region by integrating the best-fitting SED in this region (Fig.\ \ref{fig:SEDs}, top left) from $3-1100$ \micron. This yields a total infrared (TIR) luminosity of $\log{L_{TIR}}$ (erg s$^{-1}$) = 43.96, implying a SFR of 3.5 $M_\odot$ yr$^{-1}$ if we use the \citet{Kenn12} relation $\log{SFR}$ ($M_\odot$ yr$^{-1}$) = $\log{L_{TIR}}$ (erg s$^{-1}$) $-$ 43.41. This method of estimating SFR assumes that the UV photons from O and B stars are efficiently reprocessed into the infrared by the dust, and thus it does not take into account the unobscured star formation activity traced by the UV radiation or H$\alpha$ emission \citep{Bell03}. We thus consider this TIR-based estimate of SFR a lower limit to the actual SFR, and use the {\em GALEX} UV data to complete the picture.

We measure a FUV luminosity of $\log{L_{FUV}}$ (erg s$^{-1}$) = 42.19 in the stellar bar, and after accounting for dust attenuation according to $L_{FUV,\it{corr}}$ = $L_{FUV,\it{obs}} + 0.46L_{TIR}$ \citep{Hao11,Murp11}, this value becomes $\log{L_{FUV}}$ (erg s$^{-1}$) = 43.64. Using the \citet{Kenn12} relation $\log{SFR}$ ~($M_\odot$ yr$^{-1}$) = $\log{L_{FUV}}$ (erg s$^{-1}$) $-$ 43.35, this translates into an additional SFR of 1.9 $M_\odot$ yr$^{-1}$. 

The analysis based on the NUV emission produces similar results: we measure a NUV luminosity of $\log{L_{NUV}}$ (erg s$^{-1}$) = 42.35 in the stellar bar, which after accounting for dust attenuation according to $L_{NUV,\it{corr}}$ = $L_{NUV,\it{obs}} + 0.27L_{TIR}$ \citep{Hao11,Murp11}, becomes $\log{L_{NUV}}$ (erg s$^{-1}$) = 43.43. Using the \citet{Kenn12} relation, $\log{SFR}$ ~($M_\odot$ yr$^{-1}$) = $\log{L_{NUV}}$ (erg s$^{-1}$) $-$ 43.17, the NUV-based SFR is 1.8 $M_\odot$ yr$^{-1}$. We add the highest UV correction to the TIR-derived SFR to yield an upper limit of 5.4 $M_\odot$ yr$^{-1}$ on the SFR in the stellar bar of \ngc. 

\section{Discussion: Origin of the Extraplanar Dusty Filaments}
\label{sec:discussion}

The 70 \micron\ flux map and $S_{70}/S_{160}$ ratio map presented in Fig.\ \ref{fig:lines} display a filamentary morphology that is consistent with an edge-brightened biconical outflow. The filamentary morphology is expected of an evacuated cone, where the column density along the line of sight is highest along the edges of the cones due to projection effects.  Were this a filled cone, the emission would be strongest along the cone axis, becoming fainter further from this axis \citep[e.g.,][]{Veil01, Marc23}. This structure overlaps with the kpc-scale dusty filaments visible in the optical broadband ($B - R$) color maps of \citet{Phil93}, but extends to much larger scale. Edge-brightened conical structures are seen in a number of starburst- and AGN-driven outflows \citep[e.g.,][]{Veil01,Marc23}.

Before discussing the outflow scenario in more detail, we examine the possibility that the extraplanar features are instead associated with tidally-stripped material or illuminated material already existing in the halo. 
While proximity to NGC~1792 ($\sim$~130 kpc) brings support to a tidal origin for the extraplanar features, such a mechanism would not readily create four distinct tails of dust symmetrically distributed on opposite sides of the galactic plane. There are also no notable asymmetries in the HI maps of \citet{Kori93, Kori96} to suggest that the dust-bearing HI has been stripped.

Arguably the best evidence for an outflow origin for the inner filaments comes from the detection of P-Cygni Na ID $\lambda\lambda$5890, 5896 profiles with blueshifted absorption and redshifted emission in the northwestern cone  \citep{Phil93, Rupk21}, and inverse P-Cygni profiles in parts of the southwestern cone \citep{Rupk21}. In this picture, an over-pressured bubble at the center of the galaxy expands faster along the path of least resistance,  i.e.\ the minor axis of the barred galaxy disk, leading to a bicone once the wind driving the bubble reaches a few scale heights above and below the disk. The wind fluid rapidly expands through the interstellar medium (ISM), eventually 
pushing that material to the walls of a cone \citep{Chev85, Heck90, Veil05}.

In this scenario, the dust in the filaments is warmer (higher $S_{70}/S_{160}$ ratio), heated by the UV photons from young and hot stars in the central starburst
and shocks associated with the wind-ISM interface.
A simple back-of-the-envelope calculation that assumes a blackbody-like temperature dependence on luminosity, $T \sim\ {\rm 4~K}~(L/5 \times 10^{10} L_{\odot})^{1/4}~(10~{\rm kpc}/D)^{1/2}$ (where $L$ is luminosity of the heat source and $D$ is the distance between the dust and the heat source), implies that shocks must account for a considerable proportion of the dust heating. Indeed, if we treat the starburst as a point source with a luminosity $\log{(L_{FUV,cor}+L_{TIR})}$ (erg s$^{-1}$) $=$ 44.13, an overestimate of the radiation that actually heats the extraplanar dust, then dust at 10 kpc of this source of radiation will have a blackbody temperature of only $T \sim\ 4$ K, assuming the dust absorbs 100\% of that luminosity. If we instead assume that this luminosity is evenly distributed over the face of the disk, such that $F = L/\pi R^{2}$, where $R$ is the radius of the disk and the flux $F$ remains constant with height, the revised blackbody temperature becomes $T \sim\ 6~K~(L/5 \times 10^{10} L_{\odot})^{1/4}~(10~{\rm kpc}/R)^{1/2}$, yielding $T \sim\ 5$ K at 10 kpc if we take $R \sim 10$ kpc. Our $S_{70}/S_{160}$ ratio map (Fig. \ref{fig:lines}) implies extraplanar dust temperatures above 20 K at heights of 1 $-$ 10 kpc, so the extraplanar dust is likely heated by an additional source of energy as it is lifted above the disk and entrained in the outflow.  

Shocks produced at the interface between the fast wind fluid and the entrained dusty gas filaments will radiate at a rate $F_{\rm sh} = 2.28 \times 10^{-3} (V_{\rm sh}/100~{\rm km~s}^{-1})^{3.0}~(n/{\rm cm}^{-3})$ erg cm$^{-2}$ s$^{-1}$, where $F_{\rm sh}$ is the total radiative flux, $V_{\rm sh}$ is the shock velocity, and $n$ is the number density of the pre-shock material \citep{Dopi95}. This radiative flux therefore exceeds that of the starburst, $L/\pi R^{2} \approx\ 0.05$ erg cm$^{-2}$ s$^{-1}$, if $V_{\rm sh} \ga 200$ km~s$^{-1}$ and $n$ $\ga$ 1 cm$^{-3}$. Some of this radiation may contribute to the heating of the dust, although faster shocks may destroy the dust grains \citep{Dopi16}.

\begin{figure}
    \centering
    \includegraphics[width=\columnwidth]{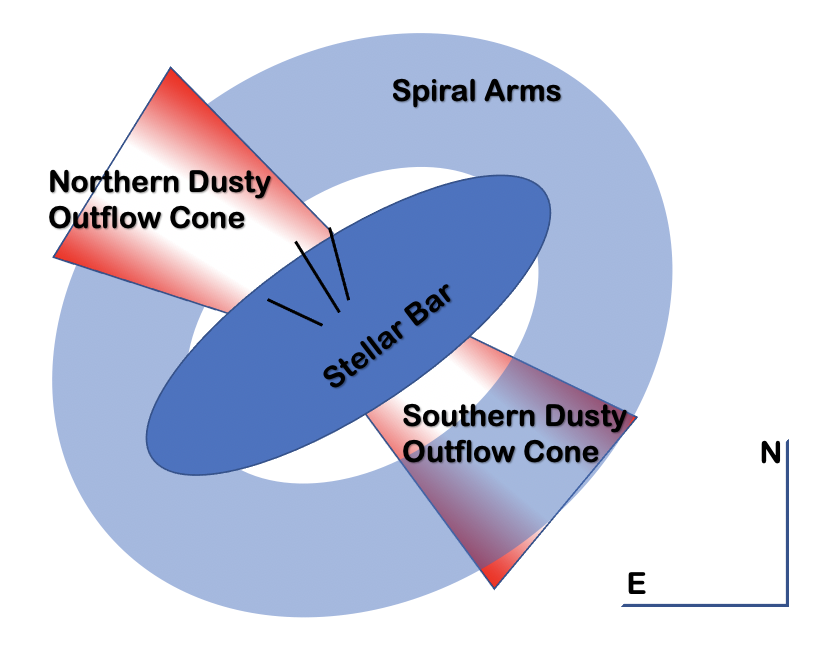}
    \caption{Geometry of the dusty biconical outflow in \ngc. The northern cone lies in front of the spiral arms, while the southern cone lies behind. The optical filaments found by \citet{Phil93} extend from the center to the northern edge of the stellar bar.}
    \label{fig:cartoon}
\end{figure}

In the end, we therefore favor an outflow origin for the extraplanar dust filaments, heated by the starburst light and radiative shocks (see Fig.\ \ref{fig:cartoon} for an illustration of the geometry). However, this interpretation requires the outflow to inject sufficient energy into the ISM on a reasonable timescale to lift the material to the observed heights. We therefore need to investigate the energetics associated with the starburst of \ngc\ and the outflow. We adopt the isothermal sheet model used by \citet{Mele15} and \citet{Veil21} to calculate the potential energy required to lift a mass out of the galactic plane: \[\Omega = 10^{52}\textrm{ ergs} \left(\frac{M}{10^5}\right)\left(\frac{z_0}{700 \textrm{ pc}}\right)^{2}\left(\frac{\rho_0}{0.185 ~M_{\odot}\textrm{ pc}^{-3}}\right)\]\[\times \text{ln}\big|\textrm{cosh}\left(\frac{z}{z_0}\right)\big|\]
where $M$ is the mass of lifted material, $z_0$ is the scale height of the galaxy (1.3 kpc, see Sec.\ \ref{sec:fluxmaps}), $\rho_0$ is the density in the midplane, and $z$ is the mean emission height. We use $\rho_0 = 0.185$ $M_\odot$ pc$^{-3}$ based on the  density of the Milky Way at the solar position \citep{Mele15}, as normal spirals like NGC 1808 tend to have similar dust properties to the Milky Way \citep{Drai07}.

In the following discussion, unless explicitly noted otherwise, we take a conservative approach and only consider the southern region between the disk and spiral arm in order to place a lower limit on the energetics required to produce the southern cone. As we see no major changes to the dust properties in the extraplanar regions (consistent temperatures), we derive flux-weighted mean heights based on the 70 \micron\ height profile. 
We find that the southern dust mass (Table \ref{tab:dust}) alone requires $1.5\times10^{54}$ ergs to be lifted to its mean height of 4.9 kpc (derived from the $S_{70}/S_{160}$ line in Fig.\ \ref{fig:ratioprofile}). Assuming a dust-to-gas ratio of 140, like that of the Milky Way \citep{Drai07,Mele15}
, the energy requirement becomes
$2.1\times10^{56}$ ergs, the equivalent energy of $\sim$ 2 $\times$ 10$^5$ Type II supernovae (SNe II), assuming each supernova supplies 10$^{51}$ ergs \citep{Murr05}. Similarly, we find that $1.9\times10^{54}$ ergs is required to lift the northern dust mass (Table \ref{tab:dust}) to its mean height of 5.1 kpc (Fig.\ \ref{fig:ratioprofile}). When the Galactic dust-to-gas ratio is incorporated, the energy requirement is 
$2.7\times10^{56}$ ergs, which could be supplied by $\sim$ 3 $\times$ 10$^5$ SNe II, or a grand total of $\sim$ 5 $\times$ 10$^5$ SNe II to produce all extraplanar filaments.

To address the kinetic energy in the outflow, we combine a model of ballistic motion, neglecting drag, with published velocity values. \citet{Rupk21} use Na I D lines to find a maximum velocity of 450 km s$^{-1}$ within the central kpc, which we can use to establish an upper limit to the kinetic energy of extraplanar dust. Given our isothermal sheet potential, material in the southern cone between $4-7$ kpc would have a flux-weighted average velocity of 320 km s$^{-1}$ if it were all initially traveling at 450 km s$^{-1}$ at 1 kpc. Material in the northern cone would have a flux-weighted average velocity of 310 km s$^{-1}$. We can account for up to another $4.2\times10^{56}$ ergs in kinetic energy, thus we expect the energy requirement to fall between $(4.8-9) \times 10^{56}$ ergs.

We use the upper limit on the velocity to constrain a minimum dynamical time for this outflow. Our ballistic model demonstrates that the gas has a nearly constant velocity while traveling from the nucleus to the edge of the 1 kpc region. Using 450 \kms\ from \citet{Rupk21}, it takes 2 Myr to travel that distance.  Traveling at 450 \kms\ at a height of 1 kpc, it requires an additional 18 Myr to reach a height of 7 kpc. We therefore estimate a total dynamical time $t_{\rm dyn}$ = 20 Myr. This timescale may be reduced if our assumption of ballistic motion is incorrect as in the case where a short-lived but highly energetic event in the past boosted the velocity of the dusty material compared with the observed kpc-scale velocity of 450 \kms. 

\begin{figure}
    \centering
    \includegraphics[width=\columnwidth]{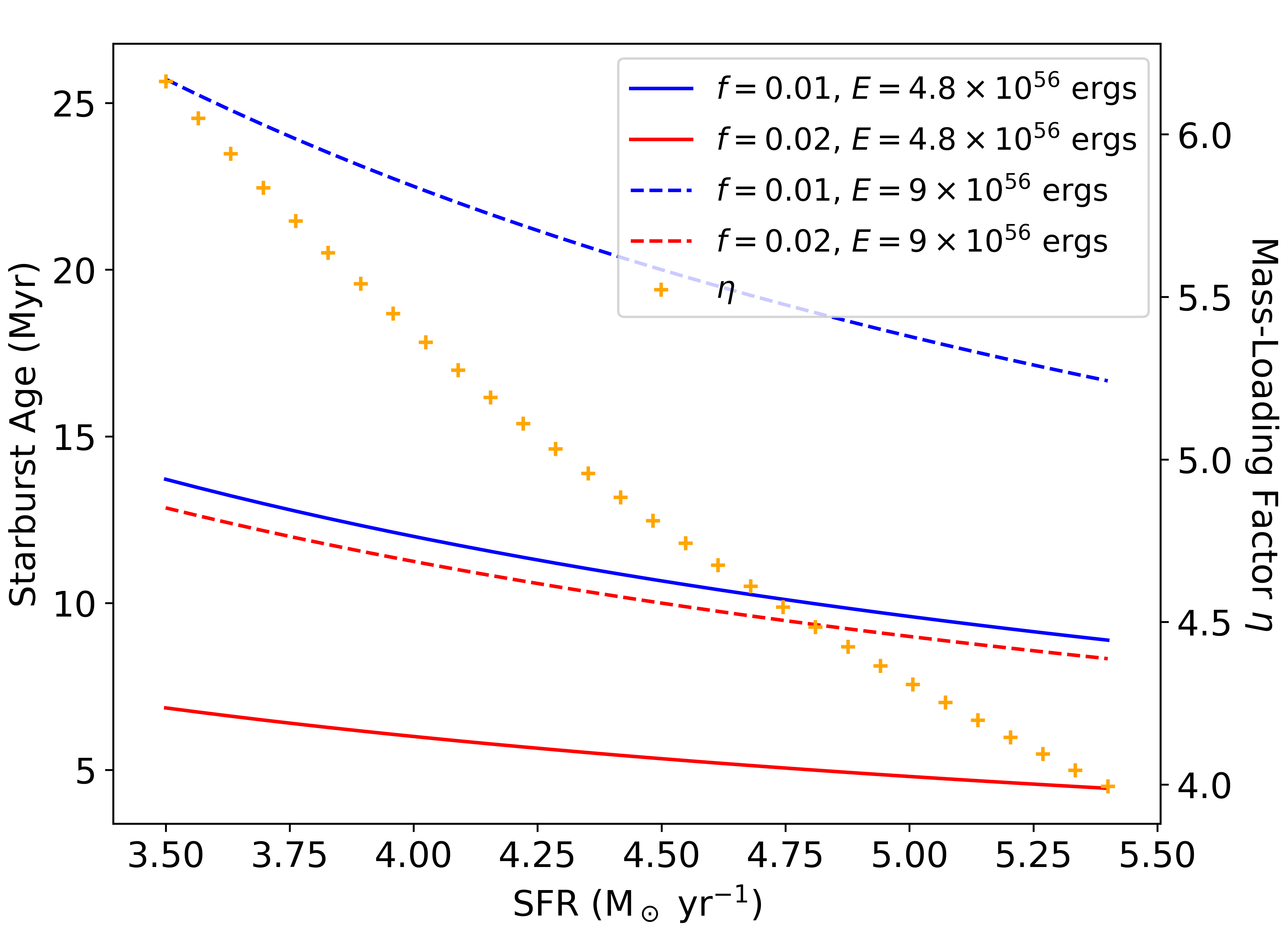}
    \caption{Minimum starburst age (left axis) needed to explain the observed extraplanar dusty material as a function of SFR, along with the corresponding mass-loading factor $\eta$ (right axis).  Solid lines trace the minimum starburst age when taking into account only the energy needed to lift the extraplanar material, while the dashed lines also consider the (maximum) kinetic energy. These numbers take into account both the northern filaments and the southern filaments within 7 kpc, using dust masses $\log(M_{d,North}/{M_\odot}) = 6.22$ and $\log(M_{d,South}/{M_\odot}) = 6.14$, and assuming Galactic dust-to-gas ratio, a MW-like central density, $\rho_0$ = 0.185 $M_\odot$ pc$^{-3}$, a dynamical time $t_{\rm dyn}$ = 20 Myr, and a 100\% energy injection efficiency, $\xi = 1$. $f$ is the number of Type II supernovae per solar mass formed, which depends on the initial mass function. Blue and red curves correspond to $f$ = 0.01 and 0.02, respectively.}
    \label{fig:timescale}
\end{figure}

Assuming a ratio of the supernova rate (SNR) to SFR of $0.01-0.02$ SNe II/$M_\odot$ (depending on the initial mass function), we find that the current star formation episode in \ngc\ is sufficiently powerful to reproduce the observed dusty cones at $\gtrsim7$ kpc as long as the current SFR  ($3.5-5.4$ $M_\odot$ yr$^{-1}$; Sec.\ \ref{sec:SFR}), and corresponding SNR ($0.03-0.11$ SNe II yr$^{-1}$), were constant for the past $\ga$ $(4-26)~\xi^{-1}$ Myr, where $\xi$ is the efficiency of energy transfer from SNe to the surrounding environment. The dependence of the minimum starburst age on the energetics is shown in Fig.\ \ref{fig:timescale}, assuming a 100\% efficiency $\xi = 1$. This efficiency is heavily dependent on the ISM properties; dense environments are able to absorb a larger portion of the  kinetic energy, but gas density and metal content can also make the ISM more susceptible to cooling. The efficiency of energy transfer from SNe to the ISM has been modeled to be as low as 10\% \citep{Thor98}, but the constraints on efficiency are somewhat relaxed by contributions from heavy stellar winds of high-mass stars which shed dust in their atmospheres. Stellar winds can supply the ISM over their lifespan with a kinetic energy comparable to a supernova \citep{Murr05}. Radiation from disk stars can also help propel an outflow, although this must be done early in the outflow history because the radiation pressure steeply drops off as $r^{-2}$.

Next, we combine our best estimates for the outflow dynamical time ($t_{\rm dyn}$ = 20 Myr) and dusty gas mass taking part in the outflow ($\log(M/M_{\odot}) = 8.63$, from Table \ref{tab:dust} and assuming $G/D \sim 140$) to derive the mass outflow rate, $\dot{M}$ = 22 $M_\odot$ yr$^{-1}$. Comparing this number with the estimated SFR ($3.5-5.4$ ~$M_\odot$ yr$^{-1}$), we get a mass-loading factor, $\eta$ $\equiv$ $\dot{M}$/SFR = $4 - 6.2$ for the outflow traced by the large-scale dusty filaments in \ngc\ (the dependence of $\eta$ on the SFR is shown in Fig.\ \ref{fig:timescale}). For comparison, \citet{Krie19} found that the nearby starburst galaxy NGC 253 has a present-day SFR between $1.7-2.8~M_\odot$ yr$^{-1}$ while ejecting mass from the central 340 pc at a rate of $14-39~M_\odot$ yr$^{-1}$ for a mass loading factor of $5.4 < \eta < 23.5$. In independent analyses of the dusty halo of NGC~891, \citet{Hodg18} and \citet{Yoon21} find that a SFR of $3.8-4.8~M_\odot$ yr$^{-1}$ over a timescale on the order 10$^{8}$ yr is capable of ejecting $\sim$ 10$^{9}$ $M_\odot$ of gas and dust, 
yielding a mass-loading factor $\eta \sim 3$. Finally, the classic starburst galaxy M82 with the iconic multi-phase galactic wind has $2 < \eta < 3$ \citep{Lero15, Veil05, Veil20, Yuan23}. The mass-loading factor we find for the dusty outflow in \ngc\ is therefore typical of starburst galaxies, given the rather large uncertainties on this estimate. For comparison with an AGN-driven outflow, \citet{Veil21} find a mass-loading factor in NGC 3079 above 100.
The discrepancy between this value and what we find in {\ngc} is in line with the conclusions of prior work that the AGN in {\ngc} is rather weak \citep{Sala18} and energetically unimportant in driving the observed large-scale outflow. 

Up to now, we have not included the energetics of the southern outflow beyond 7 kpc (light red region in Fig.\ \ref{fig:subregions}), since beyond that height the outflow emission starts to overlap with the spiral arm emission and we cannot distinguish between the two sources of emission. In that extended region where the spiral arm overlaps with the outflow, we derive a dust mass of $\log(M_d/{M_\odot}) = 6.37^{+0.11}_{-0.15}$ (Table \ref{tab:dust}) out to a height of 9 kpc, with a flux-weighted mid-height of 5.7 kpc. To lift just this dust mass out to this height requires $3.1\times10^{54}$ ergs, or $4.3\times10^{56}$ ergs when the gas is also considered.   

To reach the maximum height of 9 kpc, we must assume a higher initial velocity of 470 \kms\ for the dust $-$ material with lower initial velocity will not reach that height. We derive a dynamical time of 36 Myr for material traveling balistically from the nucleus at 470 \kms\ and reaching a height of 9 kpc with zero velocity. We expect material in the region $4-9$ kpc southwest of the bar to have a flux-weighted average velocity of 290 km s$^{-1}$. The energetics needed to produce the northern cone and {\em extended} southern cone is $0.7-1.2 \times 10^{57}$ ergs.  A constant starburst at the measured SFR can provide this energy in $(6-34)~\xi^{-1}$ Myr, where $\xi$ is the efficiency of energy injection.  Using $t_{\rm dyn}$ = 36 Myr, the mass-loading factor ranges over $2.9 < \eta < 4.5$, which is still largely consistent with those of nearby starburst-driven winds. These numbers should be considered upper limits given the likely FIR flux contamination from the spiral arm.

\section{Conclusions}
\label{sec:conclusions}
The main results from our analysis of multi-wavelength archival data on the multi-phase outflow in the starburst galaxy \ngc\ are the following: 
\begin{itemize}
    \item \her\ FIR flux maps, particularly at 70 \micron\ and 100 \micron, show extraplanar filaments that extend up to heights of $\sim$ 13 kpc from the galactic mid-plane and trace an edge-brightened bicone. Our analysis of the $S_{70}/S_{160}$ flux ratio map shows that dust along the edges of the cones is warmer than in the surrounding gas. The \spi\ $S_{8}/S_{4.5}$ flux ratio map shows emission from PAH molecules within the cones out to heights of $\sim 7$ kpc. The full extent of the southern FIR emission is uncertain because it overlaps with emission from the spiral arm in that region. 
    \item We derive the dust temperatures and masses in the stellar bar, northern emission cone, and two regions in the southern emission cone via SED fitting. Extraplanar dust constitutes at least $\sim\ 5\%$ of the total dust mass in the galaxy. 
    We derive a star formation rate SFR = 3.5 $M_\odot$ yr$^{-1}$ in the stellar bar, based on the total infrared luminosity derived from fits to the FIR data. Unobscured UV radiation may account for up to an additional 1.9 $M_\odot$ yr$^{-1}$, for a total SFR of up to 5.4 $M_\odot$ yr$^{-1}$. The starburst may not fully account for the elevated temperature of the extraplanar dust; radiative shocks may also contribute to the heating.
    \item We compare the energy needed to lift the dusty material in the biconical FIR filaments with the energy produced by the starburst in the stellar bar. Assuming a Milky-Way like dust-to-gas ratio, the energy injection rate of the current starburst must have been maintained for the past (4-26) $\xi^{-1}$ Myr, where $\xi$ is the efficiency of energy transfer from supernovae to the surrounding environment, to reproduce the dusty filaments. The range in the starburst age stems from uncertainties on the SFR and the corresponding supernova rate which depends on the initial mass function. 
    \item Combining the published velocities of the dusty neutral-atomic gas at 1 kpc with our morphological measurements of the dusty filaments, we derive an outflow dynamical time of 20 Myr, a mass outflow rate of 22 $M_\odot$ yr$^{-1}$ and mass-loading factor of $4 - 6.2$, similar to that of other nearby starburst-driven outflows.
\end{itemize} 

\vskip 0.1in

The {\em Herschel Science Archive} has proven to be a valuable source of data to map dust in nearby galaxies with galactic winds, but it does not have the required sensitivity to probe dust on CGM scale.  With the recent decommissioning of the Stratospheric Observatory For Infrared Astronomy (SOFIA), this state of affairs will persist for the foreseeable future until the launch of a next-generation far-infrared space observatory, possibly in the next decade as part of NASA's new astrophysics probe program \citep[e.g.,][]{Moul23}.

\begin{acknowledgments}

We thank Dr.\ Marcio Mel\'endez for assistance with the SED fitting, Dr.\ Alex McCormick for the initial cleaning of the \spi\ 4.5 and 8.0 \micron\ images, and Dr.\ Nadia Zakamska and Ms.\ Swetha Sankar for valuable discussions and comments that helped improve the manuscript. We also thank the anonymous referee for their constructive report that made useful suggestions, particularly regarding the heat source(s) of the extraplanar dust. S.V.\ was supported in part by NASA grants NHSC/JPL RSA 1427277, 1454738, and ADAP NNX16AF24G. 

\end{acknowledgments}

%

\vspace{5mm}
\facilities{\her, \spi, \galex, \xmm}


\software{astropy \citep{Astr13, Astr18}}





\bibliography{n1808}{}
\bibliographystyle{aasjournal}



\end{document}